\documentclass[useAMS,usenatbib]{mn2e}
\usepackage{graphicx,amsmath}

\title[]{
Jeans Instability in a Tidally Disrupted Halo Satellite Galaxy 
}
\author[]{Justin Comparetta \&  Alice C. Quillen  \\ \\ 
{Department of Physics and Astronomy, University of Rochester, Rochester, NY 14627, USA; jcompare@pas.rochester.edu} \\
}

\begin{document}
\label{firstpage}
\maketitle

\begin{abstract}
We use a hybrid test particle/N-body simulation to
integrate 4 million massless test particle
trajectories within a fully self-consistent $10^5$ particle N-body simulation.
The number of massless particles allows us to resolve fine
structure in the spatial distribution and phase space of a dwarf galaxy as 
it is disrupted in the tidal field of a Milky Way type galaxy.
The tidal tails exhibit nearly periodic clumping or a smoke-like appearance.  
By running simulations with different satellite particle
mass, halo particle mass, number of massive and massless particles and
with and without a galaxy disk,  we have determined that the
instabilities are not due to numerical noise, amplification of structure
in the halo, or shocking as the satellite passes through the disk of the
Galaxy.  We measure Jeans wavelengths and growth timescales in
the tidal tail and show that the Jeans instability is a viable explanation 
for the clumps.
We find that the instability causes velocity perturbations of
order 10 km/s.  Clumps in tidal tails present in the Milky Way could be 
seen in stellar radial velocity surveys as well as number counts.  
We find that the 
unstable wavelength growth
is sensitive to the simulated mass of dark matter halo particles. 
A simulation with a smoother halo exhibits colder and thinner tidal tails
with more closely spaced clumps 
than a simulation with more massive dark matter halo particles. 
Heating by the halo particles increases the Jeans wavelength 
in the tidal tail affecting substructure  development, suggesting
an intricate connection between   
tidal tails and dark matter halo substructure.
\end{abstract}

\section{Introduction}

There is evidence for past and ongoing accretion of small objects
by the Milky Way halo, the most dramatic object
being the disrupted Sagittarius dwarf galaxy \citep{ibata94}.
Disrupting satellites leave behind tidal tails and, 
on longer timescales, stellar streams
\citep{bekki03,helmi03,meza05,penarrubia05,purcell07,helmi08}.
Previous studies of stellar streams have used N-body simulations
to study the disruption and evolution of merging galaxies 
(e.g., \citealt{helmi06,johnston08,gomez10}).
However, most previous simulations have not placed many particles in the 
disrupting object itself.  A large number of
particles is required to resolve structure in the velocity
distribution in a small local volume 
(e.g., \citealt{minchev09,gomez10}).
After satellite disruption, satellite particles are distributed all over
the Galaxy.

Early simulations of dwarf galaxy disruption necessarily contained few particles.
The focus of many of these simulations was constraining the orientation and shape of the halo 
from observations of the Sagittarius Dwarf stream rather than searching 
for substructure in the tails themselves.
The simulations by \citet{johnston96,johnston98} 
contained only $10^4$ particles in the dwarf galaxy
and those by \citet{helmi06} only had 5000 particles.
The simulations by \citet{bullock05} had $10^4$ massive particles 
but an additional $1.2 \times 10^5$ massless test particles in their dwarf satellites.
The simulations discussed by \citet{gomez10} are described
in more detail by \citet{villalobos08} and contained    
$3 \times 10^5$ particles in the satellite.

While substructure in the form of clumps has not been detected in the Sag Dwarf streams, it has
been seen in tidal tails associated with smaller objects.
Clumped structure observed in the tidal debris of the globular cluster Palomar 5 \citep{oden02} 
is interpreted in terms of oscillations in the cluster \citep{gnedin99}
caused by a previous passage through the Galactic disk \citep{oden02,dehnen04}, epicyclic perturbations excited during tidal disruption
\citep{kupper08} and Jeans instability \citep{quillen10}.
Alternative possibilities accounting for structure in cluster tidal tails
include the effect of dark matter subhalos, as explored by 
\citet{ibata02,mayer02,johnston02,penarrubia06,valluri08,carlberg09}.

Here, we strive to carry out N-body simulations with a larger number of particles placed in the disrupting 
dwarf galaxy. We use the simulations
to look in detail at the structure of the  tidal tails during and following disruption from the dwarf galaxy.

\section{Numerical Integrations}

We first describe the N-body integrator used.  We then
describe our modifications to the integrator that allow us to
simultaneously integrate test (massless) particles.  
We then describe the initial conditions
used in our simulations and list the different simulations carried out.

\subsection{Hybrid N-body and test particle integration}

The N-body integrator used is a direct-summation code
called $\phi$GRAPE \citep{harfst07} that employs a 4th order Hermite integration
scheme with hierarchical commensurate block timesteps \citep{makino92}.
Instead of using special purpose GRAPE hardware, we use the
the Sapporo subroutine library \citep{gaburov09} that closely
matches the GRAPE-6 subroutine library \citep{makino03} but
allows the force and jerk computations to be done 
on graphics processing units (GPUs).
Force computation is done in double precision, though jerk computation is not.

We have modified the integrator so that massless particles
can be integrated simultaneously along with the massive
particles.  This is done by adding an extra parameter to 
the $\phi$GRAPE code, $N_{massive}$, 
that is the number of massive particles.
The massless and massive particles are integrated together but
only the massive particles are part of the $j$ data set used
to calculate the forces and jerks \citep{makino03}.
The GRAPE-6 subroutine library passes $n_j$ massive particles
to the GRAPE boards (or in our case to the GPU)
with the subroutine call {\it  g6\_set\_j\_particle} 
(see the GRAPE-6 User 
guide\footnote{http://www.artcompsci.org/~makino/softwares/GRAPE6/}). 
Instead of passing all particles, we pass the total number of massive
particles, $N_{massive}$, to this routine.  The accelerations
and jerks on the $i$ particles are computed by summing
terms from each of the $n_j$ particles; this is done in the GRAPE-6 
subroutines  {\it g6calc\_firsthalf} and  {\it g6calc\_lasthalf} that are
called in the gravity computation step of $\phi$GRAPE.
The number of $j$ particles in these two subroutine calls is
also changed to $N_{massive}$.  The remaining computation steps,
e.g., the predictor step, the corrector step and the identification
of active particles, remain unchanged and are done on all particles.

Computation of accelerations and jerks in an N-body simulation
require $O(N^2)$ computations for $N$ particles.
In our hybrid code $O(N_{massive}^2)$ computations are done
on the massive particles and $O(N_{massive} \times N_{massless})$
are done on the massless particles.
For example a simulation of $10^5$ massive and $10^7$ massless particles
would require $O(10^{12})$ computation steps.  This is less
than integrating $10^7$ massive particles, which would require
$O(10^{14})$ computation steps.  Thus, the hybrid scheme is a way
to integrate additional particles and better resolve
structure in phase space while not compromising the speed
of the simulation.
The hybrid scheme adopted here made it possible for us
to integrate 4 million dwarf galaxy particles during tidal disruption 
in the context of a self-consistent N-body simulation with a live halo on
a desktop computer containing off the shelf graphics cards. 
While test particle simulations can simulate this number of
trajectories without difficulty (e.g., \citealt{minchev09,quillen09}),
test particle simulations alone (not combined with N-body)
are not self-consistent.

The test-particle/N-body hybrid scheme used here is similar to the particle
cloning technique often used in celestial mechanics
integrations to improve understanding of statistical properties
of orbits  (e.g., \citealt{kaib09,masaki03}).
Test particles have been used previously to better resolve
structure in cosmological simulations.  For example,
those by \citet{bullock05} placed $1.2 \times 10^5$ test particles
in their dwarf galaxies which contained only $10^4$ massive particles.

\begin{table*}
\begin{minipage}{7.0in}
\caption{Numbers of Particles
in each Galactic Component \label{tab:tab1}}
\begin{tabular}{@{}lccccc}
\hline
\hline
Simulation No. & Disk & Bulge & Halo & Dwarf$_{massive}$ & Dwarf$_{massless}$ \\
\hline
Run 1 & 77000 & 37000 & 8000 & 9072 & $4 \times 10^6$ \\
Run 2 & 77000 & 37000 & 8000 & 60000 & $4 \times 10^6$ \\
Run 3 & 77000 & 37000 & 80000 & 60000 & $3.94 \times 10^6$ \\
Run 4 & 0 & 37000 & 80000 & 60000 & $4 \times 10^6$\\
\hline
\end{tabular}
{\\
Particle numbers for the four simulations.  Total mass in the bulge, disk and halo components is the same for all runs (except for Run 4 which does not include a disk).
When there are more particles in a given component the mass of each particle is lower. All particles in a given component have the same mass.
}
\end{minipage}
\end{table*}

\subsection{Initial conditions}

Initial particle distributions were created separately for
the Milky Way type galaxy and the dwarf galaxy.

The initial conditions for the model 
Milky Way were made with a numerical phase phase 
distribution function using the method discussed by 
\citet{widrow08} and their numerical routines which
are described by \citet{kuijken95,widrow05,widrow08}.
The code computes a gravitational potential for bulge, disk and halo
components, then computes a distribution function for each component.
N-body initial condition files are then computed for each component. 
The galactic bulge is consistent with a Sersic law for the projected
density.  The halo profile has cusp strength $\gamma$. 
The disk falls off exponentially with radius and as a sech$^2$ with height.
Parameters are those for the standard Milky Way model,
listed in Table 2 by \citet{widrow08}.  

The number of massive particles in each of the bulge, disk, halo and dwarf
components are shown in Table \ref{tab:tab1}.
The total mass of the disk is $5.31 \times 10^{10} M_\odot$, the bulge is $8.27 \times 10^9 M_\odot$, and the halo is $4.42 \times
10^{11} M_\odot$.  The halo is live.

The dwarf galaxy initial conditions were created using a King model.
The model is described by 2 parameters, a velocity dispersion, $\sigma^2$, 
and a concentration, $c$. 
The concentration, $c=r_t/r_0$,
is the ratio of the tidal radius, $r_t$, setting the outer boundary 
to the core radius, $r_0$.
Of interest is the central density $\rho_0 = {9 \sigma^2 \over 4 \pi G r_0^2 }$.
This central density sets the approximate 
location of complete tidal disruption in the background galaxy.
In Table \ref{tab:tab2} we list the properties of the dwarf galaxy. 
The total mass of the dwarf galaxy is $2.70 \times 10^8 M_{\odot}$.

The initial conditions of each simulations are identical except for the following changes.
Run 1 has a lower number of massive particles present in the dwarf galaxy than the other 3 runs,
though the total mass in the dwarf is the same.
Run 2 is our standard run.
Run 3 and 4 have 10 times more dark matter particles than the Runs 1 and 2, though  
the total halo mass is the same in all runs.
Run 4 is lacking a disk.  There is a small but insignificant change in the number of massless particles
in the dwarf galaxy in Run 3.

\subsection{Orbit}


We choose a polar orbit for the dwarf satellite.  The initial position of
the satellite is 15 kpc from the center of the Milky Way type
galaxy along the direction of the plane of the disk, and 20 kpc perpendicular to the
disc.  The velocity is 200 km/s directed toward the disk.  The orbital time is approximately
1 Gyr.  The orbit of the dwarf can be seen in the snapshots as shown in Figures \ref{fig:movie191},
\ref{fig:movie254}, \& \ref{fig:movie177}.  Each figure corresponds
to a separate run, with Figure \ref{fig:movie191} corresponding to  Run 2, 
Figure \ref{fig:movie254} corresponding to  Run 3, and Figure \ref{fig:movie177} corresponding to  Run 4.
The parameters of the King model for the satellite were adjusted so that
the satellite produced strong tidal tails but did not completely disrupt
during its first pericenter passage.

\subsection{Additional details about the simulations}

The softening length for all particles in all simulations was 0.1 kpc 
and was chosen to be the average
initial spacing between dwarf galaxy particles.  Total energy 
was conserved between 99.996$\%$ and 99.999$\%$ in all simulations.
If the chosen softening length were larger we would have failed to see 
structure on kpc scales.
If the softening length were chosen to be smaller we would have seen 
unrealistic acceleration of a small fraction
of the particles to high velocities from close approaches.

All simulations were performed for a timescale of 3 Gyr on a single 
desktop computer with two GTX 295 GPUs. Each run took between 5 and 10 days. 

\section{Results}

We first discuss the morphology of the tidal tails as seen in the different 
runs.  We compare tidal tail structure as a function of number 
of simulated massive particles in the dwarf
and number of particles in the halo.  We explore the possibility that 
the Jeans instability is the cause of the clumping
apparent in the simulations.

\begin{table}
\caption{King Satellite Model \label{tab:tab2}}
\label{tab:single}
\begin{tabular}{@{}lc}
\hline
\hline
Parameter & Value \\
\hline
$c$  & 0.672 \\
$r_0$ & 0.5 kpc \\
$\rho_0$ & $0.4 M_{\odot}/{\rm pc}^3$ \\
$\sigma$ & 25 km/s \\
\hline
\end{tabular}
{\\
Parameters for the satellite are described by a King model with parameters above.
}
\end{table}

\subsection{Morphology}

\begin{figure}
\begin{center}
\includegraphics[angle=0,width=7in]{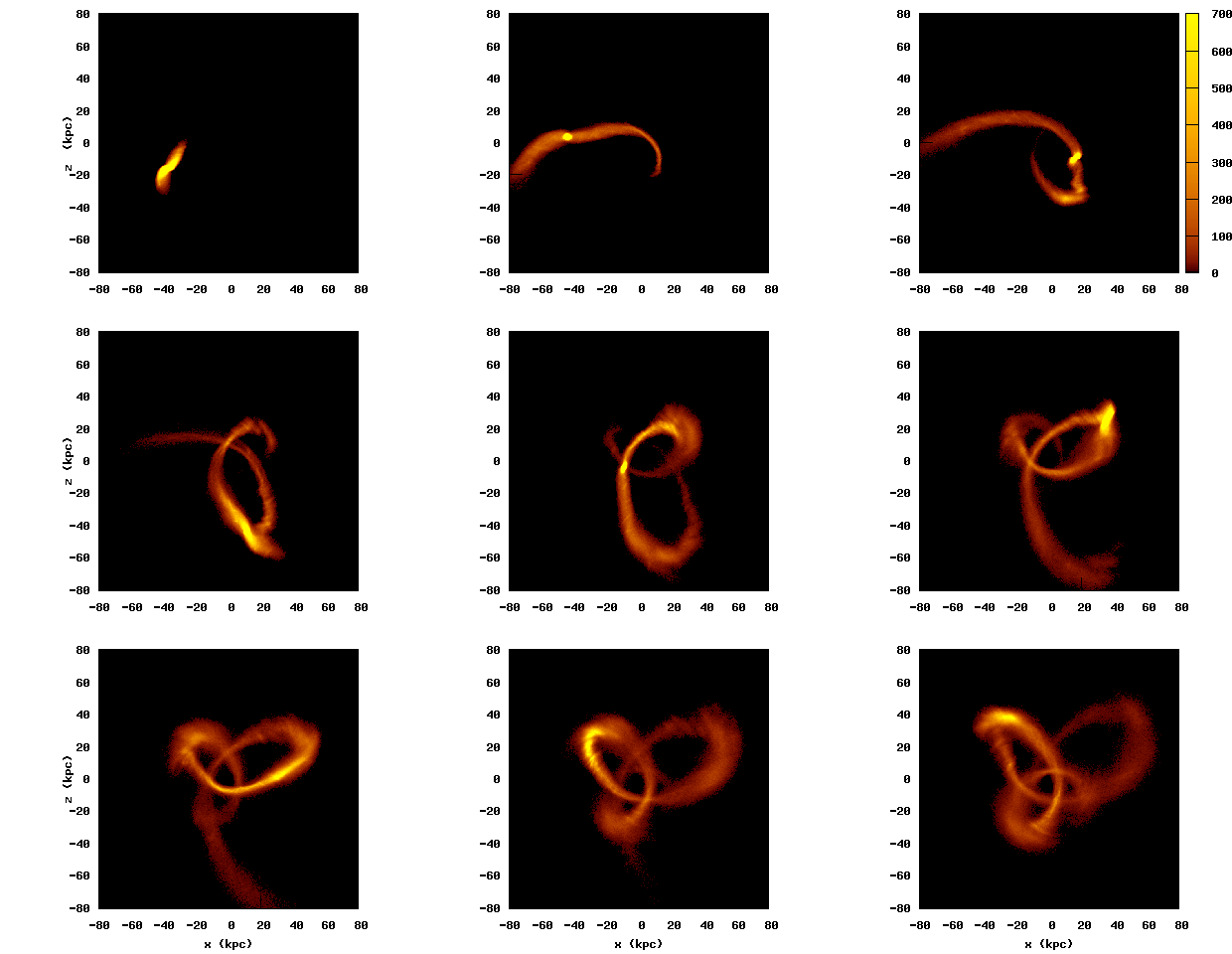}
\end{center}
\caption{
Snapshots showing projected number density of particles from
the dwarf satellite in Run 2.   
Particles are shown projected onto the $x-z$ plane.
The satellite is in a clockwise polar orbit with a Milky Way type 
galaxy centered at the origin containing a disk 
lying in the $z=0$ plane (oriented horizontally in the figure).  
Disk, halo and bulge particles are not shown.
The orbit is shown from 0.33 to 3 Gyr 
with a time of 0.33 Gyr between snapshots.
Note the development of substructure in the tidal tail which can be seen in all panels past 1.0 Gyr, and most visibly at 2.66 Gyr 
(middle panel, bottom row).  
Substructure is in the form of nearly periodically spaced density 
enhancements.  In the regions with periodic
substructure, there is no change in tail width between clump and 
interclump regions; density variations are 
ridges perpendicular to the orbit caused by longitudinal or compressive motions 
along the direction of the orbit.
\label{fig:movie191}
}
\end{figure}

\begin{figure}
\begin{center}
\includegraphics[angle=0,width=7in]{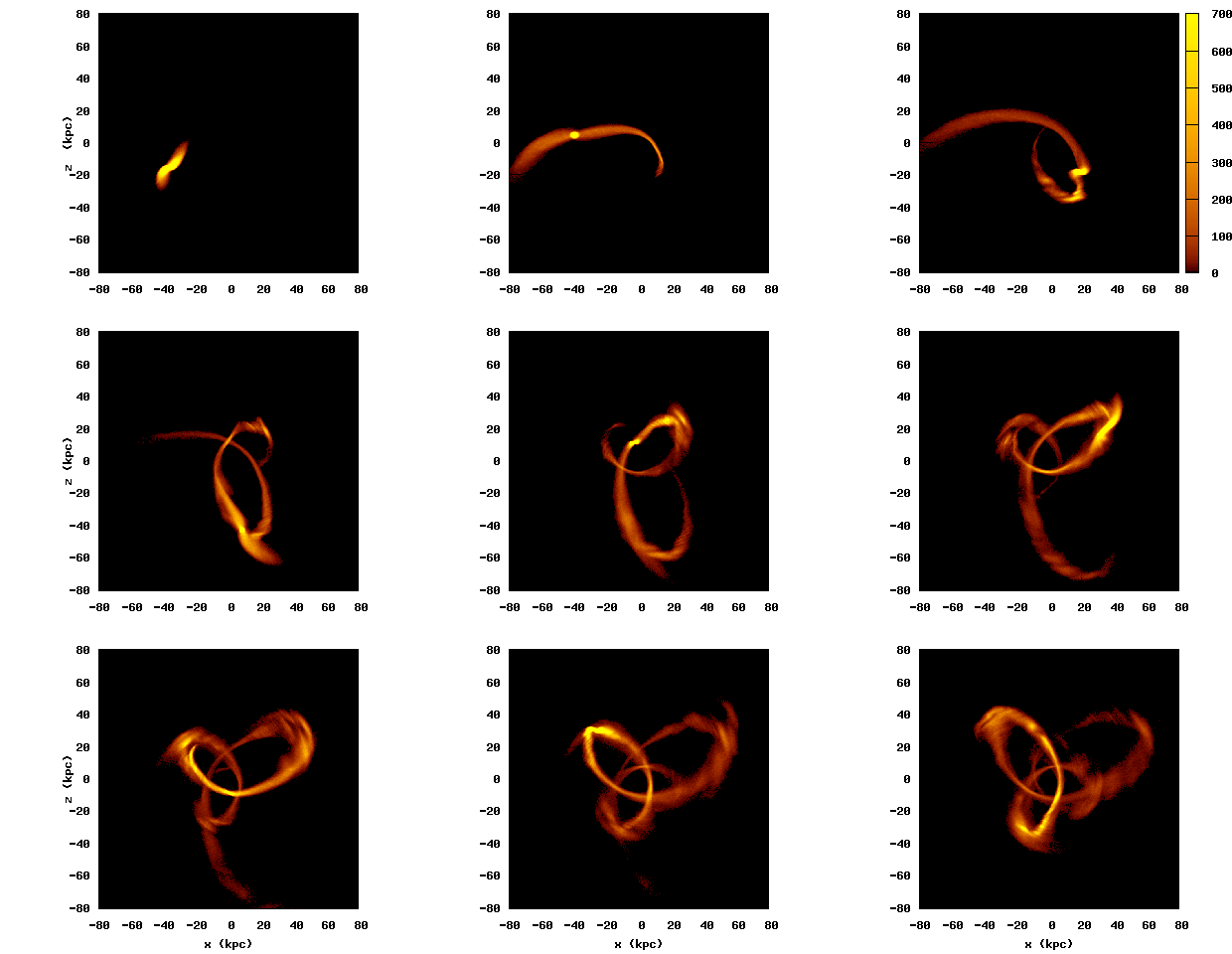}
\end{center}
\caption{
Snapshots showing projected number density of the dwarf satellite particles 
in Run 3. Similar to Figure \ref{fig:movie191}.  This simulation
has 10 times as many dark matter halo particles as that shown 
in Figure \ref{fig:movie191}; however, the orbit and mass distributions are 
identical.    Tidal tails in this simulation are narrower and denser 
than in Run 2; this suggests that the halo particles have heated
the tidal tails more in Run 2 than Run 3 due to their larger mass.  
The substructure in the tidal tails shown here
is more prominent and displays more clearly defined and closely spaced
ridges.
\label{fig:movie254}
}
\end{figure}

\begin{figure}
\begin{center}
\includegraphics[angle=0,width=7in]{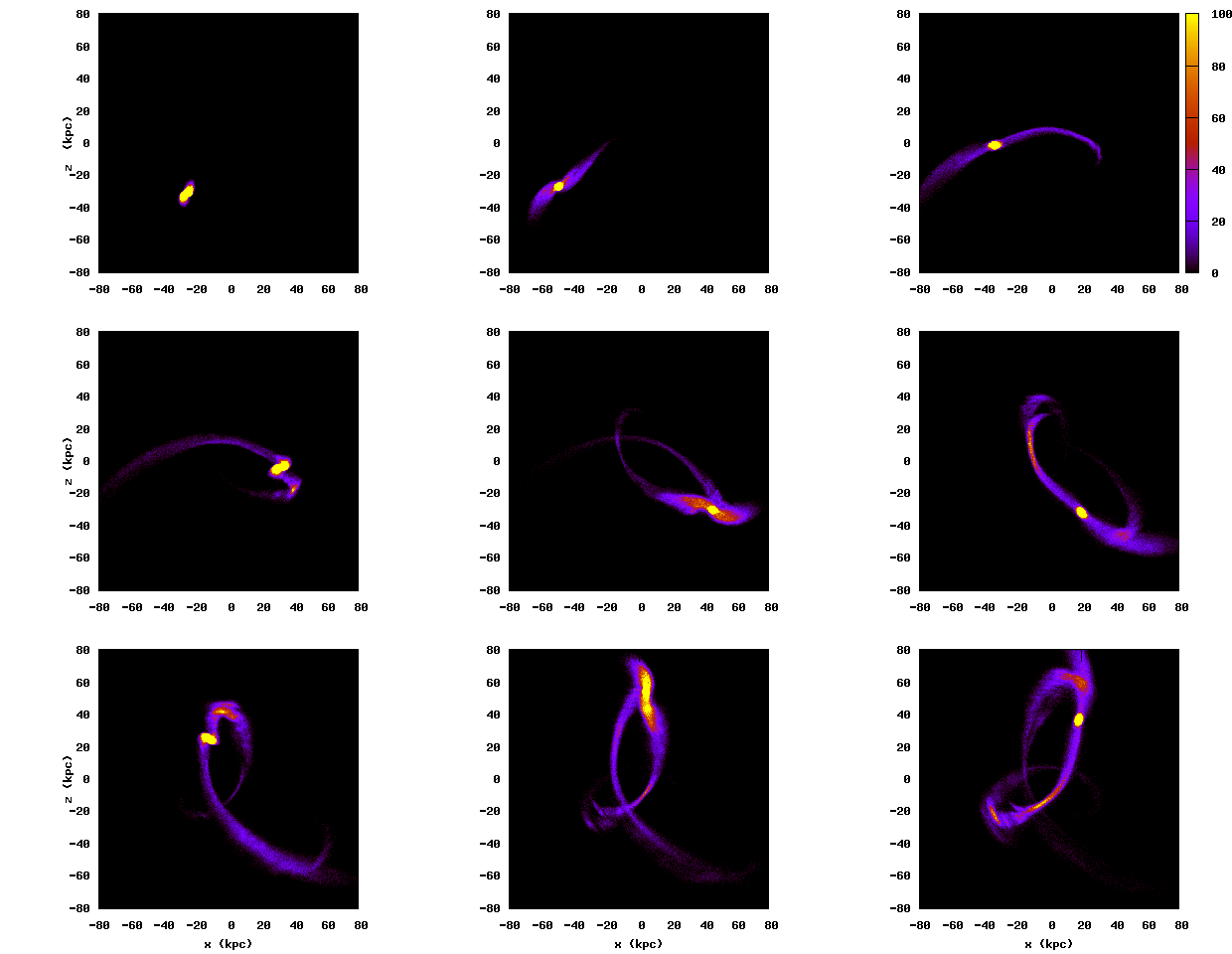}
\end{center}
\caption{
Snapshots showing projected number density of the dwarf satellite particles 
in Run 4.  Similar to Figure \ref{fig:movie191}. This simulation 
does not contain a galactic disk. However, substructure is seen in the 
tidal tails suggesting that disk shocking is not the cause
for the density enhancements seen in Runs 1--3.
\label{fig:movie177}
}
\end{figure}

Snapshots from simulation Runs 2, 3, \& 4 are shown in Figures \ref{fig:movie191},
\ref{fig:movie254}, \& \ref{fig:movie177} respectively.  
Run 1 is not shown as it is visibly indistinguishable
from Run 2.  The figures show projections into the $x-z$ plane 
of number density (histograms) in Galactocentric coordinates. 
The galaxy disk lies in the $z=0$ plane or horizontally on these plots.  
All dwarf galaxy particles are included and both massive and 
test particles are shown. 
Halo, disk and bulge particles are not shown.
Evolution of the dwarf
galaxy is shown from 0.33 Gyr to the end of the simulation at 3.0 Gyr in increments of one-third Gyr.
We saw no differences between massive and massless dwarf galaxy 
particle distributions so they are not displayed separately.

In these snapshots, it is evident that the tidally disrupted tails of the dwarf galaxy exhibit clumps, or
nearly periodically spaced density enhancements.  
When viewed sequentially as a movie this effect appears almost smoke-like in behavior.
The density enhancements can be seen in most of the snapshots past 1.0 Gyr, and more clearly in the left panels of
 Figures \ref{fig:vel191} and \ref{fig:vel254}.  
The spacing between clumps is 2-5 kpc in Run 2 but 
shorter in Run 3 and of order 1-2 kpc.
The density enhancements are in the form of nearly periodically spaced ridges that are oriented perpendicular to the orbit.

From comparisons of Run 1 to Run 2 we can test the possibility that this 
substructure is caused by heating within the tail itself and by the massive
dwarf galaxy particles in the tail.
For Run 2 we use a factor of 6.6 times more massive particles in the dwarf galaxy than in Run 1 so the mass of each dwarf particle
is 6.6 times larger in Run 1 than Run 2.  Yet the two simulations
have similar morphologies.  Noise caused by low numbers of massive particles in the dwarf is not likely responsible for the clumping.

Run 2 and Run 3 are the same except that Run 3 contains more halo particles.  The mass of each halo particle in Run 2 is $5.5 \times 10^7 M_\odot$  
whereas each halo mass in Run 3 is a tenth of this.  
Here we do see a difference between the tidal tail morphologies. 
Run 3 (with lower mass halo particles) has denser and narrower tails.  
Previous simulations have seen thickening of tidal
tails caused by heating from subhalos 
\citep{ibata02,mayer02,johnston02,penarrubia06}.  
The wider tails seen in Run 2  (Figure \ref{fig:movie191}) compared
to those Run 3 (Figure \ref{fig:movie254}) are consistent with heating caused 
by the larger masses of the halo particles in Run 2.
The mass of our halo particles in Run 2 is similar to the lowest mass subhalos simulated by \citet{ibata02}.  Their  
simulations were carried out in a smooth static background potential but included softened subhalos. 

Previous simulations have found that dark matter subhalos can also cause structure in a tidal tail \citep{valluri08,carlberg09}.  
\citet{valluri08} also used a smooth static background potential and included 
subhalos with masses in the range of
$6 \times 10^6- 6 \times 10^8 M_\odot$ as well as a smaller number of halo particles with masses up to $4.6 \times 10^{10}$.
  Clumps in the tails can be excited by nearby dark matter subhalos (see Figure 3 by \citealt{valluri08}).
Here, however, we see more prominent but also more closely spaced clumps 
in tails when the dark matter particles are of lower mass 
(Run 3; see Figure \ref{fig:movie254}).  
The larger number of clumps in Run 3 compared to Run 2 suggests that the dark matter particles
are not the cause of the substructure.
\citet{valluri08} saw clumps in the tails that were simulated in a smooth static potential and lacked subhalos (see their Figure 3 top left panel).
This suggests that if we were to carry out simulations in a smooth potential (or with more and even lower mass halo particles) that we would
continue to see clumps.

Run 4 lacks a stellar disk.   
In this simulation (see Figure \ref{fig:movie177}) 
we still see clumping in the tails.  
However, as the clumping is still 
present in Run 4 we conclude that shocking from passage through the disk is 
not the cause.  Note that the difference in the orbit shown in Figure \ref{fig:movie177} from Figures \ref{fig:movie191}
\& \ref{fig:movie254} is to be expected as the simulation is missing the potential from the galactic disk. 

In summary, our simulations show periodic clumping in the tidal tails of a disrupting dwarf satellite.   By comparing two simulations with different numbers of
massive particles in the dwarf we rule out amplification of numerical noise from massive particles in the tails as an explanation for clumps.
The simulation lacking a galactic disk also exhibits clumps, thus disk shocking cannot account for the tail structure.
The mass of the dark matter particles does affect the tail morphology, but in a way opposite to that expected. 
When the halo particles are more massive the clumps are smaller and more closely spaced rather than larger, suggesting that heating by the halo
has reduced the extent of the clumping instability.

We note that the clumps we see in the tidal tails would be  
difficult to see in a simulation containing fewer dwarf galaxy particles.
Most previous simulations may have lacked the substructure we see here
because they had fewer particles in the dwarf galaxy.
It is possible that the simulations by \citet{valluri08} do display Jeans 
instabilities (see Figures 8 and 13 which may show periodic clumps) however
it is not easy to tell as they plotted each particle individually rather 
than made histograms as we have done here.

\subsection{Jeans Instability}

Self gravity could be responsible for the substructure seen in the 
simulated tidal tails.
To test this hypothesis we measure the Jeans wavelength at different regions 
in the tails and times during the simulations.

The Jeans wavelength is 
\begin{equation}
	\lambda_J \equiv \sqrt{\pi \sigma^2 \over  G \rho_0 }.
   \label{eq:jlength}
\end{equation}
Planar perturbations on wavelengths longer than the Jeans wavelength are 
unstable and grow on a timescale
\begin{equation}
	t_{growth} \sim \left({4\pi G \rho_0}\right)^{-1/2}  \label{eq:tgrowth}
\end{equation}
whereas those wavelengths shorter than
the Jeans wavelength are damped via a process similar to Landau damping 
(e.g. Chap. 5, \citealt{BT}).

If the Jeans instability is responsible for the clumping then 
prior to clump formation we should find that the Jeans wavelength
is comparable to the distance between the clumps.  

Formally, the growth timescale is infinite at the Jeans wavelength. 
However, perturbations at all wavelengths larger than the Jeans 
wavelength are unstable. 
The wavelength of maximum growth rate
should be larger than but of order of the Jeans wavelength 
(e.g., \citealt{fridman84,BT,quillen10}).
The exact growth rate, however, is non-trivial to predict for the case
where the wavenumber of the perturbation times the width of the tail is 
on order or greater than 1, as seen in \citet{quillen10} for the regime
investigated in these simulations.
We expect that unstable perturbations with wavelength just longer than 
the Jeans wavelength are those with the fastest growth.

\subsubsection{Measuring Jeans wavelength and growth rates}

We calculate the Jeans wavelength using both massive and massless particles 
in the tidal tail.  Including only the massive 
particles results in an insufficient particle sample to compute the velocity 
dispersion in a bin small enough 
to resolve the substructure.  Since we include both massive and 
massless particle in computing the density $\rho_0$ 
in Equation \ref{eq:jlength}, 
we normalize the density by the ratio of massless plus massive to massive
particles.  We include only those particles within a half kpc of the plane 
containing the dwarf galaxy orbit (the $y=0$ plane).
In bins of $0.25 \times 0.25$ kpc in the $x,z$ directions and 1 kpc wide 
in the $y$ direction we computed sums involving both the massive 
and massless particles.
To estimate the density in the bin, we count the number of particles, multiply by the mass of the massive particles and divide by the ratio of the number 
of massless plus massive to massive particles.  
The velocity dispersion in the bin was computed from all particles in the bin and using all velocity components.
The Jeans wavelength was then computed from the mass density and velocity dispersion using equation \ref{eq:jlength}.

Growth rates are computed from the mass density using equation 
\ref{eq:tgrowth}.  The growth rate is calculated for the same regions 
that were used in computing the Jeans lengths.  
Here only the massive particles were included in the computation 
of the growth rate as we only require the density and 
we did not need to estimate the local velocity dispersion.

\begin{figure}
\begin{center}
\includegraphics[angle=0,width=7in]{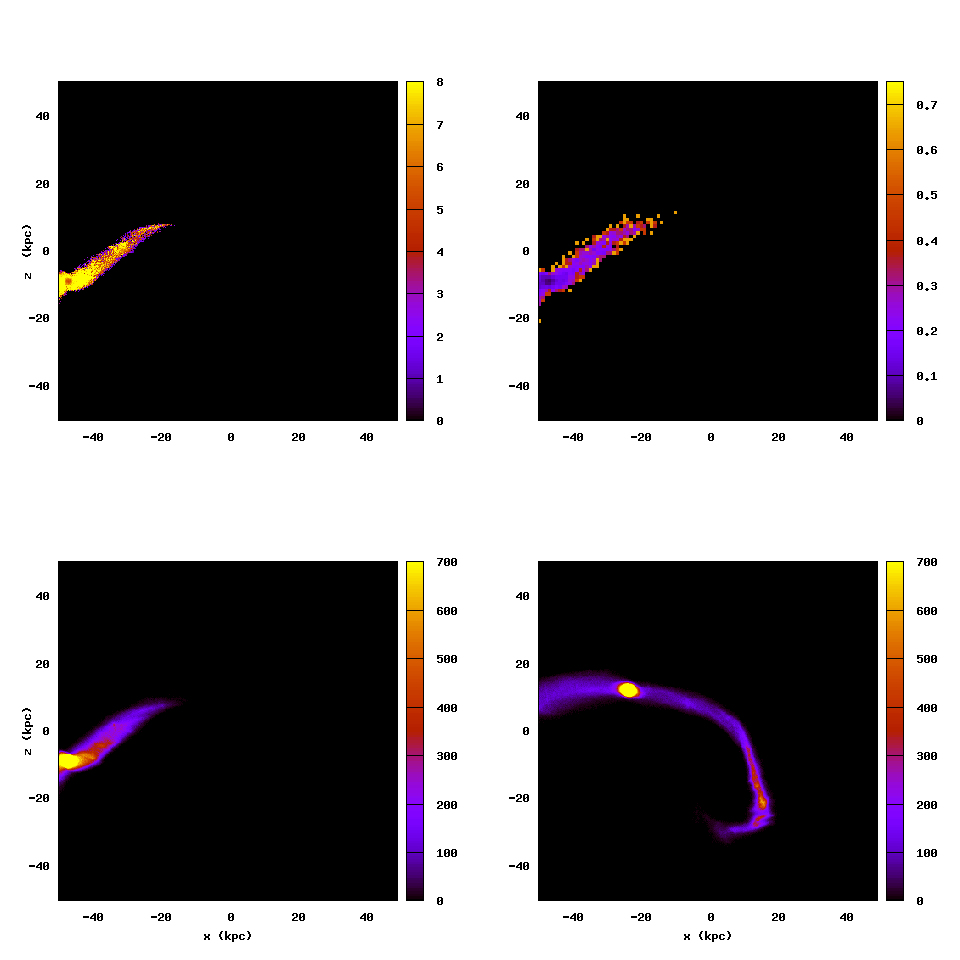}
\end{center}
\caption{Top left panel: Jean's length of the tidal tail from Run 2 
in kpc at the 0.46 Gyr timestep.  The Jeans length is computed in each
$0.25 \times 0.25$ kpc region in the $xz$ plane for all particles 
within 0.5 kpc of the $y=0$ plane containing the polar orbit 
of the satellite.  We see regions which are Jeans unstable for wavelengths 
greater than 2--3 kpc.
Top right panel:  Growth timescale in Gyr for the same region, at the
same time and in the same bins.  Growth timescales for the instability are 
about 0.2-0.3 Gyr.
Lower left panel:  Number density of the tidal tail for the same time 
and region integrated along the $y$-direction.  
Lower right panel:  Number
density of tidal tail at 0.82 Gyr which is 0.36 Gyr after 
the snapshots shown in other three panels and about 1 instability 
growth timescale later.
Substructure at the Jean's wavelength has grown in the tail 
in the expected timescale. 
\label{fig:clump46}
}
\end{figure}

\begin{figure}
\begin{center}
\includegraphics[angle=0,width=7in]{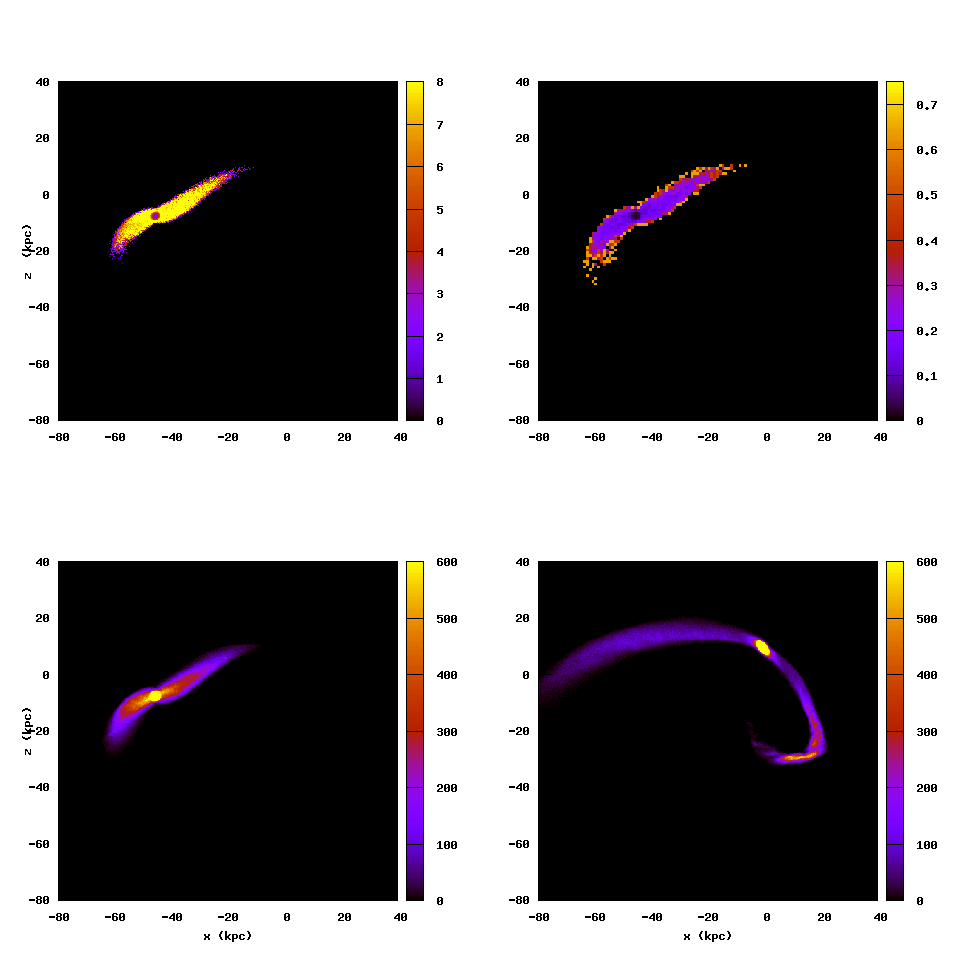}
\end{center}
\caption{Top left panel: Jeans length of the tidal tail from Run 3 in kpc at 
0.47 Gyr.  This Figure is similar to Figure (\ref{fig:clump46}). 
We see regions which are Jeans unstable for wavelengths greater than 2 kpc.
Top right panel:  Growth timescale in Gyr for tidal tail for the same 
timestep.  Growth timescales for this instability for the same region are 
about 0.3--0.4 Gyr.
Lower left panel:  Number density of tidal tail for the same timestep 
integrated along the $y$-direction.  
Lower right panel:  Number
density of tidal tail at 0.87 Gyr and 0.4 Gyr after the previous three panels.  
Structure at the Jean's wavelength has grown in the expected timescale. 
\label{fig:clump47}
}
\end{figure}

Jeans wavelengths computed in the tails at different timesteps are shown 
in Figure \ref{fig:clump46} and Figure \ref{fig:clump47}.
At the same timesteps and locations in the tail we also show the growth timescales and projected density.

In Figure \ref{fig:clump46} the top left panel shows the Jeans wavelength 
for the tidally disrupted dwarf in Run 2 at a timestep of 0.46 Gyr.   
For the upper right portion of the tail, there are sections of the tidal 
tail that have Jeans lengths of 2--3 kpc.  This is similar to 
the spacing between the clumps seen later in the simulation 
and is shown in the projected density 
in Figure \ref{fig:clump46} in the bottom right panel. 

In Figure \ref{fig:clump46} the top right panel shows the growth rate 
in Gyr for the same sections of the tidal tail used in the
computation of the Jeans length in the top left panel.   
For the same sections of the tidal tail that had Jeans lengths of 2--3 kpc, 
the growth rate is about 0.2--0.3 Gyr. 

If the observed clumping is due to the Jeans instability, we should see 
clumping become evident in the simulation after about a growth timescale.
The two bottom panels in Figure \ref{fig:clump46} show projected number 
densities of all particles in the tidally
disrupted dwarf.  The bottom left panel is a snapshot at the same time 
as the upper panels, at 0.46 Gyr, while the bottom
right panel is at 0.82 Gyr.  Clumping has formed within 0.36 Gyr as 
expected from the growth rate estimate, and the clumps
are about 3.5--4 kpc apart as would be expected from the 
Jeans wavelengths exhibited by the tail earlier in the simulation.

The same progression is illustrated for Run 3 in Figure \ref{fig:clump47} 
for the simulation with lower mass halo particles.
Here the upper right portion of the tail has a Jeans length of around 
2 kpc in the upper left panel, and a growth time scale for that region of 
0.3--0.4 Gyr is seen 
in the upper right panel.  Although no clumping is seen in the projected 
density in the lower left panel at 0.47 Gyr, clumps have formed
in the tail a growth timescale later.  These are 
shown in the bottom right panel at 0.87 Gyr 
and have a shorter spacing (compared to Figure \ref{fig:clump46}) 
as expected from the shorter Jeans wavelength 
exhibited earlier.

Within the context of first order linear perturbation theory 
(e.g., \citealt{BT}) the growth timescale for the Jeans instability
is the inverse of an exponential 
growth rate.  If perturbations present in the tail 
are small, then they would require many exponential growth timescales before 
they cause detectable density contrasts.
However, we see substructure in a time that is only on the order of 
a single growth timescale.  
There are two factors that may be contributing to this.
The original perturbations that grow may not be small,
and examination of the density at early times in the simulation 
suggests that this may be the case.
Gravitational collapse can be self-similar \citep{shu77}.
As the growth is expected to be non-linear  
the density contrast may become high on a single collapse timescale.

To summarize, we find that the spacings between the clumps 
are consistent with Jeans wavelengths
measured earlier during the simulation.  The delay timescale is similar 
to the growth timescales needed to develop the instability. 
This suggests that the Jeans instability is a viable explanation 
for the periodic substructure we see
in our simulations.

\subsection{Features in velocity space}

The periodically spaced over-densities are visible in space coordinates 
and so could be visible in stellar number counts on the sky.   
However removal of background number counts introduces noise in a 
measurement of density in a tidal tail (e.g., \citet{yanny09}).
As background number counts may be high it may be difficult 
to detect low amplitude density
density perturbations in a tidal tail from number counts alone.
Here we consider the possibility that the clumps also cause structure 
in the radial velocity field. 

\begin{figure}
\begin{center}
\includegraphics[angle=0,width=6.5in]{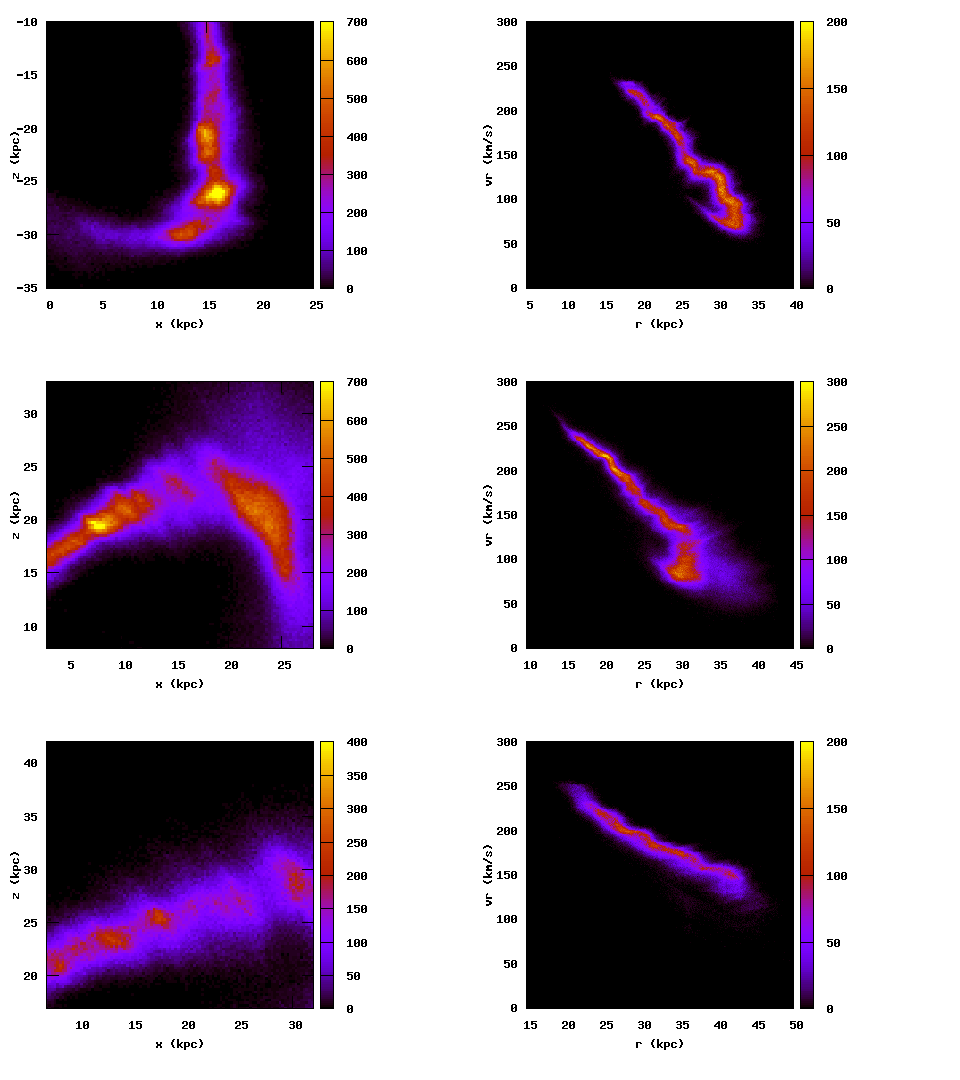}
\end{center}
\caption{Left panels:  Projected number densities at 3 different locations 
and times in Run 2 (0.86, 1.70, and 2.08 Gyr from top to bottom).  
We have chosen regions with prominent clumping in the tidal tails.  
Right panels: For particles shown on each left panel the radial velocity 
($y$-axis) is plotted against radius ($x$-axis).  
Radial velocity perturbations are seen in the density clumps.
These perturbations are of order 10 km/s and might be detectable
in a real tidal tail with high resolution spectroscopy.  
Perturbative motions are along the direction of 
orbital motion in the tidal tail. They are
are not due to wiggling or bending of the tail.  In between 
the clumps the mean velocities are divergent,  and in the clumps
they are convergent,
consistent with compressive motions due to Jeans instability.  
There are higher velocity dispersions inside the
clumps themselves implying that the development of the instability
has heated the tail. 
\label{fig:vel191}
}
\end{figure}

\begin{figure}
\begin{center}
\includegraphics[angle=0,width=7in]{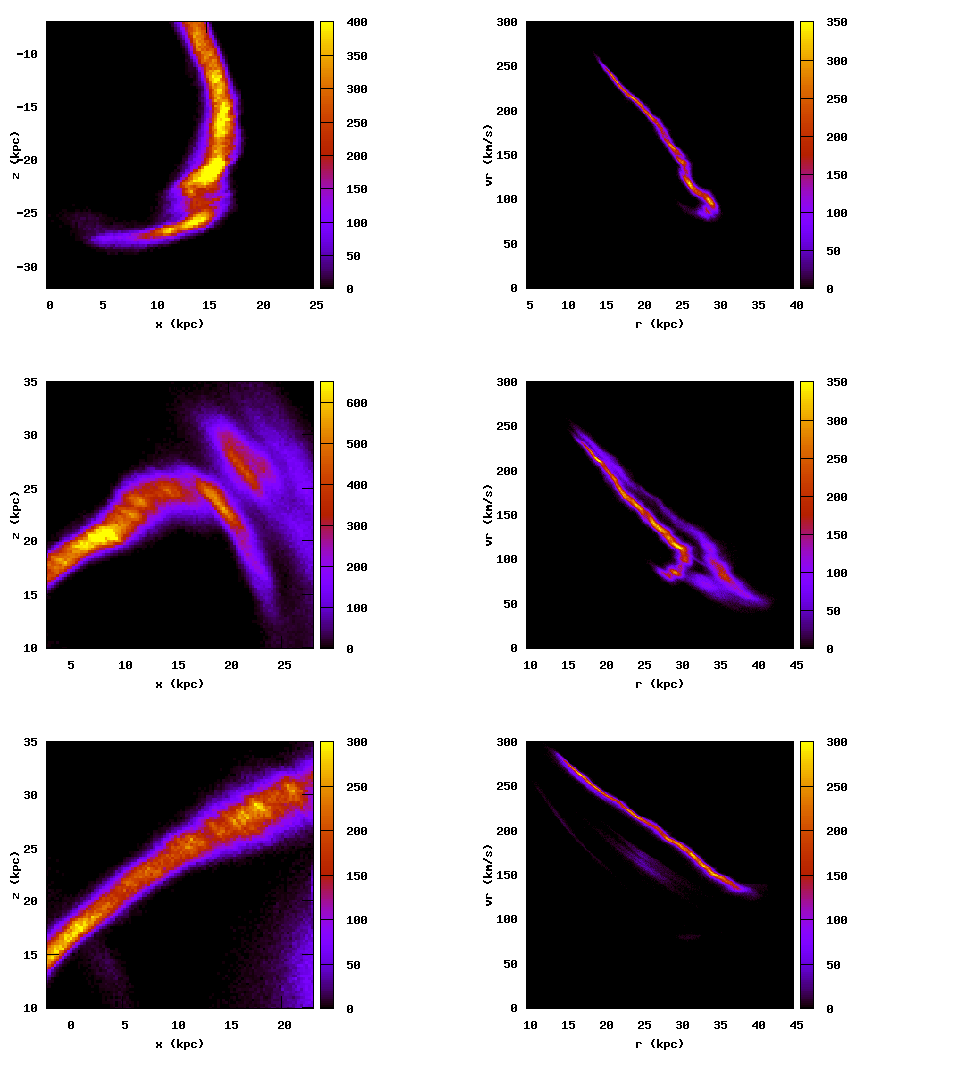}
\end{center}
\caption{ This figure is similar to Figure ~\ref{fig:vel191} but for Run 3 
and at times (0.77, 1.60, and 1.85 Gyr from top to bottom). 
Clumps are more closely spaced here than in Run 2.  
In this simulation the tails are colder and denser because of 
reduced heating by the halo.  
As a result the Jeans wavelength is smaller and shorter wavelength
perturbations have grown in the tail.
Velocity perturbations are smaller than shown in Figure \ref{fig:vel191} 
for Run 2 with larger halo particles.
\label{fig:vel254}
}
\end{figure}

In Figures \ref{fig:vel191} \& \ref{fig:vel254},
for Run 2 and Run 3, respectively, we have plotted radial velocity 
versus radius 
(right panels, in galactocentric coordinates) for several sections of 
the tidal tails exhibiting clumping (shown in number density plots in
the left panels).   
In these figures, we show on the left the projected density for small regions 
of the tidal tail.  For each panel on the left
there is a corresponding panel on the right showing the distribution 
of radial velocity $v_r$ versus galactocentric radius $r$. 
The radial velocity component would be consistent with a radial velocity
measurement by an observer near the Galactic center.   
We chose these components to roughly
illustrate phenomena that would be seen for a distant tidal 
stream as observed from the Sun.  However we did not project
components from a specific location outside the galactic center which would
have required us to specify an arbitrary 
location within the context of this simulation.

In Figures \ref{fig:vel191} \& \ref{fig:vel254},
the same regions which display substructure (clumps) in spatial coordinates 
also exhibit substructure in the velocity plots.
It is important to note that though we have plotted the velocity 
in the radial direction of a spherical galactocentric coordinate, 
we find that the 
velocity gradient points along the path of the orbit 
and is consistent with compression along the tidal tail and along the orbit.
These plots should not be misinterpreted in terms of bending of the tail.

In Figures \ref{fig:vel191} \& \ref{fig:vel254}, 
the densest regions in the left hand panels show
the density corresponds to bright vertical regions on
the $v_r$ vs $r$ plots shown on the right.  
These regions have larger ranges in the radial velocity component,
so larger velocity dispersions.   
The clumps have larger velocity dispersions than inter-clump regions.

Steps in the $v_r$ vs $r$ plots correspond to changes in the mean velocity.  
Steps are particularly visible
in the bottom right hand panel of Figure \ref{fig:vel191}.
The smooth drop in velocity with increasing radius 
in the bottom right hand panel is caused by the orbit.
If one was to subtract a smooth mean curve from this panel a sinusoidal-like oscillation would remain.  This corresponds to
positive and negative velocity perturbations about the mean orbital velocity.  
After subtraction the zeros of the sinusoidal oscillation 
lie in the inter-clump region.  These correspond to 
velocities diverging from the mean orbital velocity. 
The maxima of the sinusoidal oscillation correspond to clumps
where the velocities are converging.  
Thus, the motions are diverging in the interclump regions and converging 
in the clumps. This is the motion expected for longitudinal compressive 
motions oriented along the direction of the orbit and along the tail.

From the mass continuity equation, we can check whether the velocity 
in the tail is consistent with the growth timescales for clumping.
The mass continuity equation is
\begin{equation}
	{\partial \rho \over \partial t} + \bigtriangledown \cdot (\rho \textbf{v}) = 0.
\end{equation}
To order of magnitude this gives, 
${\Delta \rho \over \Delta t} \approx \rho {\Delta v \over \Delta x}$ 
where $\Delta x$ is the spacing between clumps, $\Delta \rho$  the difference
between clump and inter clump density,
$\Delta v$ the size of the velocity  perturbations
and $\Delta t$ the timescale of the instability.
Solving for $\Delta v$:
\begin{equation}
	\bigtriangleup v \sim {\bigtriangleup x \over \bigtriangleup t}  {\bigtriangleup \rho \over \rho}.
\end{equation}
For spacing between the clumps, $\bigtriangleup x$, of 3 kpc, growth timescales, $\bigtriangleup t$, of 0.3 Gyr, and a 
density contrast, $\bigtriangleup \rho / \rho$, of 2, we estimate 
a $\bigtriangleup v$ of 20 km/s.
This is about the size of the velocity jumps
between the clumps in the panels shown in Figures \ref{fig:vel191} \& \ref{fig:vel254}, and
implies that the velocity jumps we see in the simulation 
are consistent with compressive motions on the growth timescale 
of the instability.

We note that the clumps have larger velocity dispersion compared to 
the interclump velocity dispersions. 
This implies that the Jeans wavelength measured after the 
clumps have grown is larger than
that present prior to the growth of the instability.  When we measure
the Jeans wavelength in a region exhibiting clumps we find that it 
is much larger than the spacing
between the clumps.  This is not necessarily a contradiction as the 
current clumps grew when the tail was colder.  Not surprisingly, 
the growth of the instability itself heats the tail as gravitational energy
is converted to kinetic energy.   
However, this does present a problem for
calculating a Jeans wavelength from an observed tail that already exhibits 
the instability in the form of clumping. When we calculate the Jeans 
wavelength from the regions shown after clump formation in 
Figures \ref{fig:clump46} \& \ref{fig:clump47} 
we find that the it exceeds the distance between the clumps.
As there is an increase in the velocity dispersion caused by the instability,
it is likely that tidal tails can be thickened by the growth 
of Jeans instabilities.  
This is possibly an issue in the interpretation of heating of tails 
from dark matter substructure alone 
\citep{ibata02,johnston02,valluri08,carlberg09}.  

Previous studies have found structure in phase space in tidal tails.  
For example, phase wrapping 
has caused clumps to appear in the velocity field \citep{helmi00,minchev09}.  
However, this process
requires many orbital times to develop and appears in the velocity 
distribution after the tail has wrapped multiple times around the galaxy.
Consequently this type of phase space structure is unlikely to be 
confused with the periodic features seen in the radial velocity plots
shown in Figures \ref{fig:vel191} \& \ref{fig:vel254}.

In summary, we observe correlations between velocity dispersion, 
mean velocity and  density 
in the tidal tails consistent with a compressive instability.   
Such correlations 
might in the future be used to identify clumping via the Jeans instability and
heating from the Jeans instability to differentiate it from clumping due to 
halo substructure.

\section{Discussion}

Here we discuss differentiating between the Jeans instability and other 
mechanisms of structure formation in tidal tails.
We contrast our results with alternate explanations for clumping 
in tidal tails.  Lastly, we compare our results with 
simulations shown in \citet{valluri08}, which explore correlations 
between subhalos in the dark matter halo and 
substructure in tidal tails.

\subsection{An alternate mechanism}

An alternate explanation for the periodic over-densities in tidal tails was proposed by \citet{kupper08,kupper10}, where they attribute the
cause of the clumping to epicylic motions of the escaped stars in the tails.  
The clumps correspond 
to places where the stars in the tail slow down in their epicyclic motion. 


The analytic models explored by Kupper et al. 
require the distance to the first over-density to be 
located some multiple of the tidal radius away from the core of the disrupting 
object.  The time it takes
the first clump to form is related to the inverse of the epicyclic frequency, 
and is the time it takes stripped stars to reach the distance to
the first clump.  After twice that duration, the stars have progressed 
twice as far and form a second over-density.  Thus, in the case of a 
circular orbit and constant tidal field \citep{kupper08}, after every multiple 
of this timescale another clump forms.

There are differences between the morphology we see in our simulations
than that expected for epicyclic over-densities.
Though the clumps in our simulations 
form at intervals similar to the tidal radius of the 
initial dwarf satellite, the distance to the first clump is much 
larger than the distance between the clumps.  The periodic spacing between the
clumps is not the same as the distance to the first clump as the epicyclic 
explanation would predict.  
Whereas the strength of the clumps decreases as a function of distance from
the parent body, here we see strong clumps forming distant from the parent body.
We find that the the timing of 
the creation of the clumps in our simulations differ from that expected from 
an epicyclic explanation.  Clumps in a region grow 
simultaneously, not in order of distance from the dwarf galaxy core.
While epicyclic over-densities are a promising explanation for clumps
seen in Palomar 5's tail they are unlikely to be the explanation for
the clumps seen in the simulations presented here.

\subsection{Halo substructure}

Clumping in tidal debris has previously been investigated by \citet{valluri08},
 who found that sub-halos were responsible
for increased substructure in tidal tails.  
This can be seen in their sky projections in Figure 3 comparing 
star particles in 
simulations with and without the presence of subhalos.  
A smooth halo  (top left panel of their Figure 3) produces less clumpy
debris than a halo with substructure (top right panel).  
Similar to our simulations, they also find the the substructure is observable
in velocity space, as demonstrated in their Figure 6 showing radial 
velocity vs radius of their star particles, again in simulations
with and without subhalos.

While most previous works (e.g.,
\citet{ibata02,mayer02,johnston02,penarrubia06,carlberg09}) 
found that halo substructure heats tidal tails, \citet{valluri08} found in
some cases that the tails in smooth halos were colder and denser than
in halos with more substructure.  It is possible that Jeans instability 
at shorter wavelengths
was responsible for additional heating in these simulations. If so
the connection between tidal tail and halo substructure may be more complex
than previously considered.

\section{Summary and Conclusion}

We have used hybrid test particle/N-body integrations to
increase the number of particles integrated within an N-body
integrator so that we can 
more accurately resolve substructure.
We have used our code to study the tidal disruption of a dwarf galaxy
in a polar orbit of a Milky Way type galaxy.  We have placed additional 
test particles
in the dwarf galaxy so that we can resolve fine structure in the tidal tails.

In our simulations, we have found that stellar tidal tails can exhibit periodic 
ridges oriented perpendicular
to the orbit.  Such structure had not previously been noticed in similar N-body 
simulations, possibly
because the tidal tails did not contain sufficient numbers of particles.  
However similar structure has
been seen in simulations of tidal tails that include gas \citep{wetzstein07}
and in this setting the clumps are interpreted in terms of a
gravitational instability.

By comparing simulations with different numbers of massive dwarf and 
halo particles and with and without a disk,
we have considered several explanations for the formation of clumps.  
We have ruled out the following methods for the formation 
of periodically spaced over-densities
in our simulation: shocking by the galactic disk, 
numerical noise associated with under-populating 
the dwarf galaxy, amplification of structure in the halo, and epicyclic motions in the tidal tails.

We have measured the Jeans wavelength prior to the growth of the 
substructure and found that the tails are unstable to Jeans instability. 
The wavelengths of subsequently formed clumps are approximately consistent 
with the Jeans wavelength measured in the tail prior to formation.
The timescale for growth is approximately consistent with the estimated 
growth timescale.  These estimates suggest that Jeans instability is
a viable interpretation of the clumps exhibited by our simulations.
We find that the spacing between clumps is sensitive to the mass of
our simulated dark matter halo particles.  This is likely because
heating by dark matter 
particles can increase the Jeans wavelength of the tidal tail.

We find that the clumps are also visible in a radial velocity
projections suggesting that 
Jeans instabilities may be observable in tidal tails in our galaxy
not only number counts but also in 
phase space using comparisons of radial velocity versus distance or
position on the sky.

In the future, we expect increasingly
rich data samples expanding the number of stars in the Milky 
Way with measured properties.  These surveys may make it possible
to probe for or rule out substructure in tidal tails such as exhibited by
our simulations.   If Jeans instabilities occur in tidal tails then associated heating caused by them should not be interpreted in terms of heating by 
halo substructure alone. Furthermore the fastest growing unstable wavelength 
may be sensitive to heating from the dark matter substructure implying that
there may be a complex connection between halo and tidal tail substructure.

The simulations carried out here were made using a modest direct N-body
code on a desktop computer.
Future studies could test the results presented here with
a more sophisticated N-body code (such as a tree code),
and integrate more particles by doing the simulations on a supercomputer.
Further studies can be carried out in different mass and stripping regimes for 
the disrupting object, such as globular clusters.  
We would also be interested in carrying out simulations that would
more fully probe the possible connection between dark matter substructure and
tidal tail morphology by simulating a more detailed halo than 
considered here.

\vskip 2.0truein
We thank Larry Widrow for giving us and helping us with his code 
GalacticICS. We thank Jeff Bailin, Chris Purcell and Heidi Newberg 
for helpful communications.
The King model was generated using code made available\footnote{http://www.physics.mcmaster.ca/~syam/software.html}
by Sergey Mashchenko.  
We thank Evghenii Gaburov and Stefan Harfst 
for making $\phi$GRAPE and Sapporo available. 
Support for this work was provided by NSF through award AST-0907841.

{}

\clearpage

\clearpage

\clearpage

\clearpage

\clearpage

\clearpage

\clearpage

\end{document}